\documentclass{article}
\usepackage{amsthm,enumitem,mathtools,simpler-wick}
\usepackage{amsmath,amsfonts,amssymb,slashed,color}
  \usepackage{soul} 
\renewcommand\hl[1]{#1}
\usepackage{authblk}

\usepackage{caption}
\usepackage{subcaption}

\usepackage{feynmp-auto}

\def\comma{,}
\def\cs{\mathrm{cs}}

\def\dirac{D}
\newcommand{\diraccom}[1]{[D \comma #1]}

\DeclareMathOperator{\tr}{Tr}

\newcommand{\br}[1]{\langle#1\rangle}
\newcommand{\dbr}[1]{\llangle#1\rrangle}

\makeatletter
\DeclareMathOperator{\exterior}{\@ifnextchar^\@exterior{\@exterior^{}}}
\def\@exterior^#1{\mathop{\bigwedge\nolimits^{\!#1}}}
\makeatother

\makeatletter
\newsavebox{\@brx}
\newcommand{\llangle}[1][]{\savebox{\@brx}{\(\m@th{#1\langle}\)}%
  \mathopen{\copy\@brx\kern-0.5\wd\@brx\usebox{\@brx}}}
\newcommand{\rrangle}[1][]{\savebox{\@brx}{\(\m@th{#1\rangle}\)}%
  \mathclose{\copy\@brx\kern-0.5\wd\@brx\usebox{\@brx}}}
\makeatother

\def\red{\textcolor{red}}
\def\blue{\textcolor{blue}}
\def\green{\textcolor{green}}

\begin{document}

\unitlength = 1mm

\title{One-loop corrections to the spectral action}

\author{Teun D.H. van Nuland}

\author{Walter D. van Suijlekom}

\affil{\small Institute for Mathematics, Astrophysics and Particle Physics\\
  Radboud University Nijmegen, Heyendaalseweg 135,\\
	6525 AJ Nijmegen, The Netherlands\\\medskip \texttt{t.vannuland@math.ru.nl, waltervs@math.ru.nl}}


\maketitle

\begin{abstract}
  We analyze the perturbative quantization of the spectral action in noncommutative geometry and establish its one-loop renormalizability in a generalized sense, while staying within the spectral framework of noncommutative geometry. Our result is based on the perturbative expansion of the spectral action in terms of higher Yang--Mills and Chern--Simons forms. In the spirit of random noncommutative geometries, we consider the path integral over matrix fluctuations around a fixed noncommutative gauge background and show that the corresponding one-loop counterterms are of the same form so that they can be safely subtracted from the spectral action. A crucial role will be played by the appropriate Ward identities, allowing for a fully spectral formulation of the quantum theory at one loop. 
\end{abstract}


\begin{fmffile}{graphs}

  \fmfset{wiggly_len}{2mm}

  \fmfset{dot_len}{1.5mm}

  \section{Introduction}
  Noncommutative geometry \cite{C94} offers a spectral viewpoint to geometry that allows to simultaneously capture field theories and gravity in a single framework. In fact, it allows for a unified geometrical derivation of the Standard Model of particle physics minimally coupled to gravity \cite{CCM07,Sui14}, including the Higgs mechanism and the see-saw mechanism to yield masses for the right-handed neutrinos. This extends beyond the Standard Model to yield Pati--Salam grand unification \cite{CCS13b,CCS15}, which is currently one of the few candidate BSM-theories that is still found to be compatible with experiment. Variations on particle theories obtained in the same framework are considered in \cite{BroS11,DLM14,DLM14b,DM14,BF14,BBS16,DS18,DAS18,BS20,BS20b}, while the more foundational aspects on quanta of geometry were considered in \cite{CCM15}. 

  The key ingredient in this description of field theories arising from noncommutative spaces is the spectral action principle \cite{CC96}. It yields Lagrangians that are based solely on the spectrum of a given Dirac operator on a noncommutative spacetime. In the applications to particle physics phenomenology one then adopts the usual renormalization group methods to arrive at couplings and mass parameters at lower energy. Even though the appearance of such experimentally testable results from a geometrical framework valid at high-energies is very intriguing, we must confess that this step is a weak point of the noncommutative approach to particle physics. Indeed, it means that in the passage to the quantum theory one looses the elegant spectral and unifying picture that one started with and which one admired so much.

  In this paper, we take a crucial step in the quantization program and analyze the form of loop corrections to the spectral action. Working in a very general context, in fact beyond \cite{CC06,CIS20}, we find that the resulting quantum fluctuations can be entirely formulated within the same unifying spectral framework and is thus a major improvement with respect to the usual RG-approach to the spectral action. The approach we take to the perturbative quantization of the spectral action is that  of random noncommutative geometries \cite{GS19b,AK19,KP20} (see also \cite{BG16,GS19a} for computer simulations). More specifically, we adopt the background field method for which the path integral will be defined over all matrix fluctuations around a fixed noncommutative gauge background.

  The key mathematical input is given by our paper \cite{NS21}, which gives a perturbative expansion of the spectral action in terms of noncommutative integrals over higher Yang--Mills and Chern--Simons forms.
  We will here show that the one-loop corrections to the spectral action are of exactly the same form, and can thus safely be subtracted as counterterms from the spectral action. \hl{This establishes one-loop renormalizablity in the generalized sense of \mbox{\cite{GW96}}, where one allows for infinitely many counterterms. }


\section{Diagrammatic expansion of the spectral action}
\label{sect:sa}
The spectral action \cite{CC96} is defined on the eigenvalue spectrum $\{\lambda_k\}_k$ of a Dirac operator $\dirac$ by
$$
\tr f(\dirac) = \sum_{k} f(\lambda_k)
$$
for some suitable even function $f$. We want to analyze the spectral action for perturbations $\dirac \to \dirac+V $ by bosonic gauge fields of the form $V =a_j \diraccom{b_j}$ (summation over $j$ understood), \hl{where $a_j,b_j$ are coordinate functions on a noncommutative space}. Even though our analysis is valid in the general setting of noncommutative geometry \cite{C94,C96} the most interesting cases that occur in physics are:
\begin{itemize}
\item Hermitian matrix models where both $\dirac$ and $V$ are hermitian matrices.  
\item Almost-commutative geometries $M \times F$, where $M$ is the spacetime manifold with Dirac operator $\slashed{\partial}$ and $F$ is a discrete noncommutative space describing the internal degrees of freedom, also equipped with a `finite' Dirac operator $D_F$. The gauge fields $V$ describe both Yang--Mills gauge fields $\mathbb{A}$ and scalar (Higgs)  fields $\Phi$ in the sense that
  $$
V= \underbrace{ a_j \slashed \partial (b_j)}_{\slashed{\mathbb{A}}} + \gamma_5 \underbrace{ a_j [D_F,b_j]}_{\Phi},
$$
More details, also on the applications to particle physics, can be found in \cite{CCM07,CCS13b,DLM14,Sui14,CIS20}. 
  \end{itemize}

Our starting point is the following expansion of the spectral action \cite{Sui11,Skr13,NS21}:
\begin{equation}
  \label{eq:exp}
  S_\dirac[V]:=
\tr\left( f(\dirac+V) - f(\dirac) \right) = \sum_{n=1}^\infty \frac 1 n \langle \underbrace{V,\ldots,V}_{n}\rangle.
\end{equation}
The brackets stand for the following contour integrals:
$$
\langle V_1, \ldots, V_n \rangle \\=  \tr  \oint \frac{dz}{2\pi i} f'(z)  V_1 (z-\dirac)^{-1}   \cdots V_n (z-\dirac)^{-1}
$$
where $V_1,\ldots, V_n$ are gauge fields as above; this can be represented nicely as a Feynman diagram:
\vspace{1mm}
\begin{equation}
  \langle V_1, \ldots, V_n \rangle = 
 \scalebox{.7}{ \parbox{6cm}{
    \hspace{1cm}
    \begin{fmfgraph*}(30,30)
    \fmfset{arrow_len}{3mm}
    \fmfsurroundn{e}{6}
    \fmflabel{$V_1$}{e4}
    \fmflabel{$V_2$}{e3}
    \fmflabel{$V_3$}{e2}
    \fmflabel{$V_4$}{e1}
        \fmflabel{$V_n$}{e5}
        \fmf{photon}{e1,i1}
        \fmf{photon}{e2,i2}
        \fmf{photon}{e3,i3}
        \fmf{photon}{e4,i4}
        \fmf{phantom}{e6,i6}
        \fmf{photon}{e5,i5}
        \fmfrcyclen{fermion,left=0.25,tension=2}{i}{6}
        \fmf{phantom,left=.5,tension=0.3}{e1,e5}
        \fmfipath{p}
\fmfiset{p}{vpath(__e1,__e5)}
        \fmfi{dots}{subpath (length(p)/4,3*length(p)/4) of p}
  \end{fmfgraph*}}}
  \label{eq:bracket}
\end{equation}


The loop diagram nicely reflects the cyclicity of the bracket: $\br{V_1,\ldots,V_n}=\br{V_n,V_1,\ldots,V_{n-1}}$. The second crucial property is that
\begin{equation*}
\br{a V_1,\ldots,,V_n}-\br{V_1,\ldots,V_n a}
        =\br{\diraccom{a}, V_1,\ldots,V_n}
\end{equation*}
\hl{for any (noncommutative) coordinate function $a$.}
In fact, this identity boils down to the following {\em Ward identity}, 
$$
(z-\dirac)^{-1} a - a (z-\dirac)^{-1} = (z-\dirac)^{-1} \diraccom {a} (z-\dirac)^{-1},
$$
and may be represented diagrammatically:
\begin{equation}
   \label{eq:ward}
\parbox{1.5cm}{    \begin{fmfgraph*}(10,10)
    \fmfset{arrow_len}{3mm}
      \fmfleft{lb,l}
      \fmfright{rb,r}
      \fmf{fermion}{l,r}
      \fmf{dashes}{r,rb}
                        \fmfv{label=$a$,label.angle=-90}{rb}
\end{fmfgraph*}}
   - \quad 
 \parbox{1.5cm}{       \begin{fmfgraph*}(10,10)
    \fmfset{arrow_len}{3mm}
\fmfright{rb,r}
     \fmfleft{lb,l}
            \fmf{fermion}{l,r}
            \fmf{dashes}{lb,l}
                  \fmfv{label=$a$,label.angle=-90}{lb}
   \end{fmfgraph*}}=
  \parbox{1.5cm}{
      \begin{fmfgraph*}(15,10)
    \fmfset{arrow_len}{3mm}
        \fmfleft{lb,l}
       \fmfright{rb,r}
       \fmfbottom{cb}
       \fmf{fermion}{l,c,r}
       \fmffreeze
       \fmf{photon}{cb,c}
       \fmflabel{$\diraccom{a} $}{cb}
                  \end{fmfgraph*}}
\end{equation}

\subsection{The brackets as noncommutative integrals}
We want to express the amplitudes corresponding to the above loop diagrams in terms of suitable noncommutative integrals \cite{C94} ({\em cf. } \cite[Eq. (4.182)]{CM07} or \cite[Sect. 4.2]{NS21}. They are defined by 
\\[2mm]
\begin{align}
  \label{eq:bracket-cochain}
  \int_{\phi_n} a^0 da^1 \cdots da^n 
  &:= \scalebox{.7}{ \parbox{6cm}{
    \hspace{1.5cm}
    \begin{fmfgraph*}(25,25)
    \fmfset{arrow_len}{3mm}
    \fmfsurroundn{e}{6}
    \fmflabel{$a^0\diraccom{ a^1}$}{e4}
    \fmflabel{$\diraccom{ a^2}$}{e3}
    \fmflabel{$\diraccom{ a^3}$}{e2}
    \fmflabel{$\diraccom{ a^4}$}{e1}
        \fmflabel{$\diraccom{ a^n}$}{e5}
        \fmf{photon}{e1,i1}
        \fmf{photon}{e2,i2}
        \fmf{photon}{e3,i3}
        \fmf{photon}{e4,i4}
        \fmf{phantom}{e6,i6}
        \fmf{photon}{e5,i5}
        \fmfrcyclen{fermion,left=0.25,tension=2}{i}{6}
           \fmf{phantom,left=.5,tension=0.3}{e1,e5}
        \fmfipath{p}
\fmfiset{p}{vpath(__e1,__e5)}
        \fmfi{dots}{subpath (length(p)/4,3*length(p)/4) of p}
  \end{fmfgraph*}}}.
\end{align}

\medskip

\noindent For example, for one external edge we find

\begin{align}
\br{V} =   \br{a_j \diraccom{b_j}} &=%
 \scalebox{.7}{  \parbox{2  cm}{
    \begin{fmfgraph*}(20,20)
    \fmfset{arrow_len}{3mm}
      \fmfleft{l}
      \fmfright{r}
      \fmf{photon,tension=.5}{l,c}
      \fmfv{label=$a_j\diraccom{b_j}$,label.angle=-90}{l}
      \fmf{phantom,tension=.1}{r,c}
      \fmf{fermion,left,tension=.5}{r,c,r}
   \end{fmfgraph*}}}
  = \int_{\phi_1} A,
\end{align}
where we have defined $A= a_j d b_j$ as the {\em universal} gauge form underlying the physical gauge field $V = a_j \diraccom{b_j}$. Note that the vanishing of this tadpole diagram corresponds to the vanishing of the first derivation of the spectral action under perturbations $D \mapsto D+V$. For natural choices of $D$ one may thus expect this term to vanish and, in fact, \cite{CC06} works under this `vanishing tadpole' assumption.

For two external edges, we apply the Ward identity \eqref{eq:ward} and derive
\begin{align*}
  \br{V,V}&=
  \scalebox{.7}{\parbox{6cm}{
    \hspace{1.3cm}
  \begin{fmfgraph*}(15,15)
    \fmfleft{l}
    \fmfright{r1,r2}
    \fmf{photon}{l,c1}
    \fmfset{arrow_len}{3mm}
    \fmf{fermion,left=1,tension=.5}{c1,c2,c1}
    \fmf{dashes}{r2,c2}
    \fmf{photon}{r1,c2}
    \fmfv{label=$a_j\diraccom{ b_j}$,label.angle=-120}{l}
    \fmflabel{$\diraccom{ b_{j'}}$}{r1}
    \fmflabel{$a_{j'}$}{r2}
   \end{fmfgraph*}} }\\[8mm]
  &=\qquad \scalebox{.7}{\parbox{2.7cm}{
     \begin{fmfgraph*}(15,15)
  \fmfright{r}
    \fmfleft{l1,l2}
    \fmf{photon}{r,c1}
    \fmfset{arrow_len}{3mm}
    \fmf{fermion,left=1,tension=.5}{c1,c2,c1}
    \fmf{dashes}{l2,c2}
    \fmf{photon}{l1,c2}
    \fmflabel{$\diraccom{ b_{j'}}$}{r}
    \fmflabel{$a_j\diraccom{ b_j}$}{l1}
    \fmflabel{$a_{j'}$}{l2}
  \end{fmfgraph*}}}
  +  \qquad \scalebox{.7}{\parbox{3cm}{
    \begin{fmfgraph*}(15,15)
    \fmfsurroundn{e}{3}
    \fmflabel{$a_j\diraccom{ b_j}$}{e3}
    \fmflabel{$\diraccom{ a_{j'}}$}{e2}
    \fmflabel{$\diraccom{ b_{j'}}$}{e1}
        \fmf{photon}{e1,i1}
        \fmf{photon}{e2,i2}
        \fmf{photon}{e3,i3}
        \fmfset{arrow_len}{3mm}
        \fmfrcyclen{fermion,left=0.5,tension=.5}{i}{3}
  \end{fmfgraph*}}}
  \\[6mm]
  &= \int_{\phi_2} A^2 
  + \int_{\phi_3} A d A.
\end{align*}
Similarly, by applying the Ward identity several times one finds that \cite{NS21} 
\begin{align*}
  \br{V,V,V}&=\int_{\phi_3}A^3+\int_{\phi_4}AdAA +  \cdots 
  ,\\
	\br{V,V,V,V}&=\int_{\phi_4}A^4+ \cdots . 
\end{align*}
We now introduce a noncommutative integral $\int_\psi$ that differs from $\int_\phi$ by a total derivative:
\begin{equation}
  \label{eq:psi}
\int_{\psi_{2k-1}}  \omega = 
\int_{\phi_{2k-1}}  \omega - \frac 1 2 \int_{\phi_{2k}}  d \omega
\end{equation}
and rewrite the above brackets in terms of $\psi_1$ and $\psi_3$, as well as the remaining $\phi_2$ and $\phi_4$.
For the first two terms, we readily find
$$
\int_{\phi_1}A + \frac 12 \int_{\phi_2}A^2 = 
\int_{\psi_1}A + \frac 12\int_{\phi_2} (dA +A^2).
$$
while after a slightly more involved derivation we also find for the next few terms that
\begin{align*}
&  \frac 12 \int_{\phi_3}AdA+ \frac 1 3 \int_{\phi_3}A^3+ \frac 1 3 \int_{\phi_4}AdAA + \frac 1 4 \int_{\phi_4}A^4
  \\
  &\quad= \frac 12 \int_{\psi_3} \left ( AdA+ \frac 2 3 A^3 \right) + \frac 14 \int_{\phi_4} (dA+A^2)^2+ \cdots .
\end{align*}

The important message from the above derivation is that the expansion of the spectral action yields Yang--Mills and Chern--Simons terms. In fact, if we write $F = dA+A^2$ for the curvature and 
  \begin{gather*}
             \cs_1(A) = A; \qquad  \cs_3(A) = \frac 12 \left( A dA + \frac 2 3 A^3 \right),
             \end{gather*}
then it turns out that the expansion has the following form of a Yang--Mills--Chern--Simons theory:
\begin{align*}
S_D[V] &= \int_{\psi_1} \cs_1(A) + \frac 12 \int_{\phi_2} F \\
&\qquad + \int_{\psi_3} \cs_3(A) +\frac 1 4 \int_{\phi_4} F^2
+ \cdots 
\end{align*}
Quite surprisingly, the systematics behind this derivation persists at all orders \cite{NS21}, while being based solely on the cyclicity of the loop diagram and the Ward identity \eqref{eq:ward}. It yields the following expansion for the spectral action
\begin{equation}
  \label{eq:YmCs}
S_D[V] = \sum_{k=1}^\infty \left( \int_{\psi_{2k-1}}  \cs_{2k-1} (A) +\frac 1 {2k} \int_{\phi_{2k}}  F^{k} \right).
\end{equation}
The higher-order {\em Chern--Simons forms} are defined as in \cite[Section 11.5.2]{Nak90} by 
           \begin{equation}\label{eq:cs}
\cs_{2k-1}(A) := \int_0^1 A (t dA + t^ 2 A^2)^{k-1} dt.
           \end{equation}
Again based solely on cyclicity of the loop diagram and the Ward identity, one can show that the integrals over $\phi_{2k}$ and $\psi_{2k-1}$ define even and odd cyclic cocycles, respectively; we refer to \cite{NS21} for more details.

\section{Loop corrections to the spectral action}
In order to analyze the quantum theory corresponding to the above classical action functional $S_D[V]$ we adopt the background field method. That is to say, we take the background fields to be gauge fields of the form $V =a_j \diraccom{b_j}$. However, the path integral is defined over the ensemble of all finite-size hermitian complex-valued matrices. This is in the spirit of random noncommutative geometries in the sense of \cite{GS19b,AK19,KP20} (see \cite{BG16,GS19a} for computer simulations). As in these works, we consider the dimension, say $N$, of these matrices as a regularizing cutoff of our model, which should eventually be sent to $\infty$, while allowing us to realize our quantum theory as a hermitian matrix model. 

In fact, for such finite-size matrices $Q =(Q_{kl})$, the brackets can be conveniently expressed in terms of divided differences of $f'$ \cite{Sui11}: 
\begin{align*}
\frac 12   \langle Q, Q\rangle &= \frac 12\sum_{k,l} Q_{kl} Q_{lk} f'[\lambda_k,\lambda_l]\\
\frac 13   \langle Q, Q,Q \rangle &= \frac 13 \sum_{k,l,m} Q_{kl} Q_{lm} Q_{mk} f'[\lambda_k,\lambda_l, \lambda_m]
\end{align*}
{\em et cetera}, where $\lambda_k$ are the eigenvalues of $D$. Recall that the first two divided differences are defined by $f'[x,y] = (f'(x)-f'(y))/(x-y)$ and $f'[x,y,z]= (f'[x,y]-f'[y,z])/(x-z)$.

\begin{figure}

  \begin{subfigure}[t]{.45\linewidth}
    \includegraphics[scale=.45]{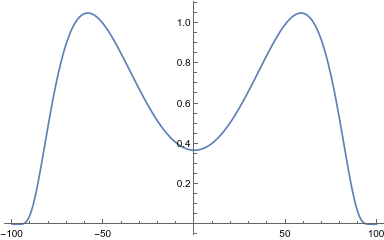}
    \caption{An example of a positive function: $f(x) = (1+a x^2) \Phi(b x)$ with $\Phi$ a bump function and $ a = 1/900, b = 1/100 $.}
    \label{fig:bumppoly}
  \end{subfigure}\hspace{5mm}
  \begin{subfigure}[t]{.45\linewidth}
  \includegraphics[scale=.3]{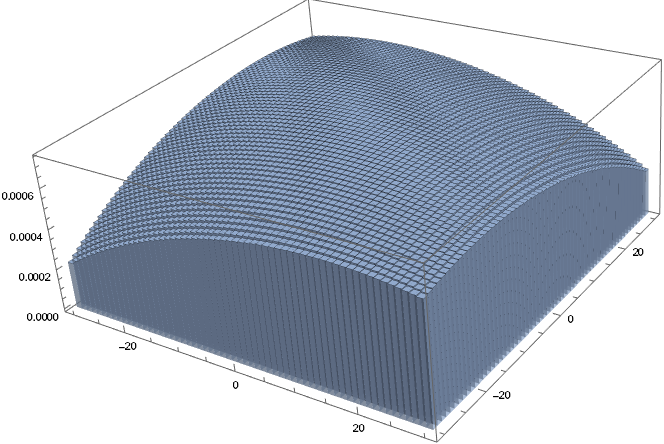}
  \caption{The divided difference $f'[\lambda_k,\lambda_l]$ for this function $f$.}
  \end{subfigure}

  \caption{The inverse gauge propagator $f'[\lambda_k,\lambda_l]$ for the $N=61$ smallest eigenvalues of the Dirac operator on the circle ({\em i.e.} $ \lambda_k,\lambda_l = -30, -29, \ldots, 30$).}
  \label{fig:diffdiv-T2}
  \end{figure}

We now make the assumption that the first divided difference of $f'$ is strictly positive on the $N$ relevant eigenvalues of $D$ (see Figure \ref{fig:diffdiv-T2}). We may then perform the Gaussian integration \hl{as in\mbox{ \cite[Section 2]{BIZ80}}, without the need for introducing a gauge-fixing and ghost sector,} to get for the propagator:
$$
\wick{\c Q_{kl} \c Q_{mn} }=\frac{ \int Q_{kl}  Q_{mn} e^{-\frac 12 \br{Q,Q}} dQ} {\int e^{-\frac 12 \br{Q,Q} } dQ} = 
\delta_{kn}\delta_{lm} G_{kl}
$$
in terms of $G_{kl} := \frac 1 {f'[\lambda_k, \lambda_l]}$. Notice that the inverse propagator is bounded, which is in stark contrast to the usual unbounded nature of inverse propagators in ordinary local quantum field theory. We see this as another manifestation of the regularizing properties of the spectral action, in line with \cite{Sui11b,ILV11,KLV13,ASZ15}.

It is an interesting problem to analyze the form of the propagator for more general $f$, including a possible gauge fixing, for instance along the lines of \cite{IS17,Ise19} or by means of orthogonal polynomials as in \cite{BIZ80}. 



In any case, we are now in a position to consider higher-loop contributions to the spectral action, and, in particular, all one-particle irreducible $n$-point Feynman graphs. Their (possibly divergent) amplitudes form the starting point of the renormalization process of the spectral action.


\subsection{Ward identity for the gauge propagator}
In addition to the Ward identity \eqref{eq:ward} for the fermion propagator, we claim that we also have the following Ward identity for the gauge propagator:


\begin{equation}
  \label{eq:ward-gauge}
 \parbox{2.1cm}{
  \hspace{.5cm}
    \begin{fmfgraph*}(25,20)
       \fmfset{arrow_len}{2mm}
      \fmfleft{l1,l2}
      \fmfright{r1,r2}
      \fmf{fermion,left=.2,tension=.5}{l2,v2,l1}
      \fmf{photon}{v2,v3}
            \fmf{fermion,left=.2,tension=.5}{r1,v3,r2}
            \fmffreeze
            \fmftop{b}
                              \fmf{dashes}{v3,b}
                              \fmflabel{$a$}{b}
\end{fmfgraph*}}
 - \parbox{2.2cm}{
   \begin{fmfgraph*}(25,20)
       \fmfset{arrow_len}{2mm}
      \fmfleft{l1,l2}
      \fmfright{r1,r2}
      \fmf{fermion,left=.2,tension=.5}{l2,v2,l1}
      \fmf{photon}{v2,v3}
            \fmf{fermion,left=.2,tension=.5}{r1,v3,r2}
            \fmffreeze
                  \fmftop{b}
                  \fmf{dashes}{v2,b}
                  \fmflabel{$a$}{b}
\end{fmfgraph*}}
=\parbox{2.2cm}{
    \begin{fmfgraph*}(25,20)
       \fmfset{arrow_len}{2mm}
      \fmfleft{l1,l2}
      \fmfright{r1,r2}
      \fmf{fermion,left=.2,tension=.5}{l2,v2,l1}
          \fmftop{b}
          \fmf{photon}{b,bc}
             \fmflabel{$\diraccom{a}$}{b}
      \fmf{photon}{v2,c1}
      \fmf{fermion,left=1,tension=.5}{c2,c1}
      \fmf{fermion,left=.5}{c1,bc,c2}
      \fmf{photon}{c2,v3}
            \fmf{fermion,left=.2,tension=.5}{r1,v3,r2}
      \end{fmfgraph*}}
\end{equation}
where every fermion loop adds a minus sign. Indeed, the left-hand side is
  \begin{align*}
    \wick{ \c Q_{ik} \c Q_{lm} a_{mn} }-   \wick{ a_{im} \c  Q_{mk}  \c Q_{ln}}   & =  G_{ik} \delta_{im} \delta_{kl} a_{mn} - G_{ln} \delta_{mn} \delta_{kl} a_{im} \\
    &= ( G_{ik}- G_{nk} )\delta_{kl} a_{in}
\end{align*}   
   while for the right-hand side we use the defining property of the divided differences to find:
\begin{align*}
  -   \wick{ \c Q_{ik} \c Q_{rp}} a_{pq} (\lambda_p -\lambda_q)&\wick{ \c Q_{qr} \c  Q_{ln}} f'[\lambda_p, \lambda_q, \lambda_r]\\
    &= - G_{ik} \delta_{ip} \delta_{kr} G_{qr} \delta_{qn} \delta_{rl} a_{pq} (\lambda_p -\lambda_q)f'[\lambda_p, \lambda_q, \lambda_r]\\
    & =  G_{ik} G_{nk} \left( f'[\lambda_k, \lambda_n] -  f'[\lambda_i, \lambda_k] \right)  \delta_{kl} a_{in}.
    \end{align*}
The two expressions coincide because of the very fact that the free propagator is the inverse of the divided difference.

\subsection{Two-point functions at one-loop}
The two-point graphs at one-loop 
are given in Table \ref{table:1loop-2pt}. The external fields $V_1,V_2$ should be assigned to the external legs in all different cyclical manners.

The amplitude for the first graph is given by
\vspace{-4mm}
\begin{align}
  \label{eq:2ptA}
 \scalebox{.7}{\parbox{2.5cm}{
     \begin{fmfgraph*}(25,25)
       \fmfset{arrow_len}{2mm}
       \fmfleft{l}
       \fmftop{r1}
       \fmfbottom{r2}
       \fmf{photon,tension=.4,label=$V_1$}{l,v1}
       \fmf{fermion,left=.5,tension=.2}{v1,v2,v3,v1}
       \fmf{phantom,tension=.1}{v2,r1}
       \fmf{phantom,tension=.1}{v3,r2}
       \fmfright{r}
       \fmf{photon,tension=.3,label=$V_2$}{r,w1}
       \fmf{fermion,left=.5,tension=.2}{w1,w3,w2,w1}
       \fmf{phantom,tension=.1}{w2,r1}
       \fmf{phantom,tension=.1}{w3,r2}
       \fmf{phantom,tension=.05}{v2,w2}
       \fmf{phantom,tension=.05}{v3,w3}
       \fmfset{wiggly_len}{2mm}
       \fmf{photon,tension=.1}{v2,w3}
       \fmf{photon,tension=.1}{v3,w2}
  \end{fmfgraph*}}} &= \sum_{\begin{smallmatrix} i,j,k \\ l,m,n\end{smallmatrix}} (V_1)_{ij} \wick{ \c1 Q_{jk} \c2 Q_{ki} (V_2)_{lm} \c1 Q_{mn} \c2 Q_{nl} }  f'[\lambda_i,\lambda_j,\lambda_k]  f'[\lambda_l,\lambda_m,\lambda_n]  \nonumber \\
 &= \sum_{i,k} (V_1)_{ii} (V_2)_{kk} G_{ik}^2 f'[\lambda_i, \lambda_i,\lambda_k]f'[\lambda_i, \lambda_k,\lambda_k] .
  \end{align}
In particular, there is no running loop index in this expression and so this diagram remains finite even when the size $N$ of the matrices is sent to $\infty$. We conclude that the amplitude of this graph is not relevant for renormalization purposes.

\begin{figure}
    \begin{subfigure}[b]{.45\linewidth}
        \includegraphics[scale=.4]{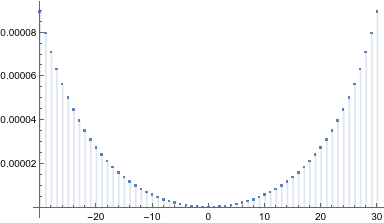}
        \caption{Summands in \eqref{eq:vertexcontr} for $(\lambda_i,\lambda_j) = (0,0)$}
      \end{subfigure}
      \hfill
      \begin{subfigure}[b]{.45\linewidth}
  \includegraphics[scale=.4]{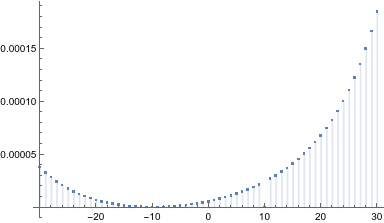}
        \caption{Summands in \eqref{eq:vertexcontr} for $(\lambda_i,\lambda_j) = (10,0)$}
      \end{subfigure}

  \begin{subfigure}[b]{.45\linewidth}
        \includegraphics[scale=.4]{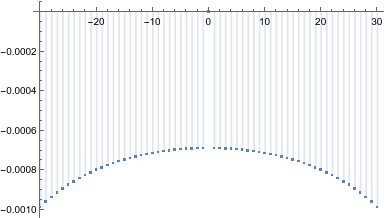}
        \caption{Summands in \eqref{eq:vertexcontr2} for $(\lambda_i,\lambda_j) = (0,0)$}
      \end{subfigure}
      \hfill
      \begin{subfigure}[b]{.45\linewidth}
  \includegraphics[scale=.4]{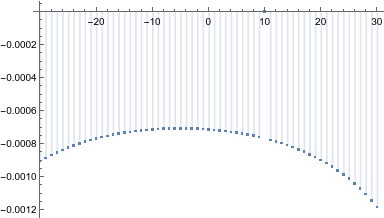}
        \caption{Summands in \eqref{eq:vertexcontr2} for $(\lambda_i,\lambda_j) = (10,0)$}
      \end{subfigure}

      \caption{The behaviour of the summands (indexed by $\lambda_k$ running from $-30$ to $30$) for the vertex contribution  in \eqref{eq:vertexcontr} and \eqref{eq:vertexcontr2} 
        for the Dirac operator on the circle and function $f$ as in Figure \ref{fig:bumppoly}.}
  \label{fig:divergence}
  \end{figure}

We then turn to the second graph in Table \ref{table:1loop-2pt}, and compute
\begin{align}
 \scalebox{.7}{ \parbox{2.5cm}{
    \begin{fmfgraph*}(25,25)
       \fmfset{arrow_len}{2mm}
       \fmfset{wiggly_len}{2mm}
       \fmfleft{l}
       \fmftop{r1}
       \fmfbottom{r2}
       \fmf{photon,tension=.4,label=$V_1$}{l,v1}
       \fmf{fermion,left=.5,tension=.2}{v1,v2,v3,v1}
       \fmf{phantom,tension=.1}{v2,r1}
       \fmf{phantom,tension=.1}{v3,r2}
       \fmfright{r}
       \fmf{photon,tension=.3,label=$V_2$}{r,w1}
       \fmf{fermion,left=.5,tension=.2}{w1,w3,w2,w1}
       \fmf{phantom,tension=.1}{w2,r1}
       \fmf{phantom,tension=.1}{w3,r2}
       \fmf{photon,tension=.2}{v2,w2}
       \fmf{photon,tension=.2}{v3,w3}
  \end{fmfgraph*}}}&= \sum_{\begin{smallmatrix} i,j,k \\ l,m,n\end{smallmatrix}} (V_1)_{ij} \wick{ \c1 Q_{jk} \c2 Q_{ki} (V_2)_{lm} \c2 Q_{mn} \c1 Q_{nl} } f'[\lambda_i,\lambda_j,\lambda_k]  f'[\lambda_l,\lambda_m,\lambda_n]   \nonumber \\
 &= \sum_{i,j,k} (V_1)_{ij} (V_2)_{ji} G_{ik}G_{kj} f'[\lambda_i, \lambda_j,\lambda_k]^2.
\label{eq:vertexcontr}
\end{align}
We find that this amplitude has a potential divergence in the limit that $N \to \infty$ (see Figure \ref{fig:divergence} for the behaviour of the summands). As such it should be subtracted from the effective action in order to render the theory finite after removal of the regulator. 

For the final diagram with two external lines we compute its amplitude to be:
\vspace{-4mm}
\begin{align}
 \scalebox{.7}{  \parbox{3cm}{
    \begin{fmfgraph*}(30,30)
       \fmfset{arrow_len}{2mm}
       \fmfleft{l1}
       \fmfright{l2}
       \fmftop{r1,r2}
       \fmf{photon,tension=1,label=$V_1$}{l1,v1}
       \fmf{photon,tension=1,label=$V_2$}{l2,v2}
       \fmf{fermion,left=.5,tension=2}{v1,v4,v3,v2,v1}
       \fmf{phantom,tension=1}{v4,r1}
       \fmf{phantom,tension=1}{v3,r2}
      \fmf{photon,left,tension=.05}{v4,v3}
 \end{fmfgraph*}}}&= \sum_{ i,j,k,l} (V_1)_{ij} \wick{ \c Q_{jk} \c Q_{kl} } (V_2)_{li}\label{eq:vertexcontr2}
 f'[\lambda_i,\lambda_j,\lambda_k,\lambda_l] \\ \nonumber 
 &= \sum_{i,j,k} (V_1)_{ij} (V_2)_{ji} G_{jk} f'[\lambda_i, \lambda_j,\lambda_j,\lambda_k].\nonumber
\end{align}
Again, this graph amplitude is potentially divergent in the limit $N \to \infty$ and should thus be subtracted. The same applies to the same graph but with $V_1$ and $V_2$ exchanged.

\begin{table}
    \begin{tabular}{p{.3\linewidth}p{.3\linewidth}p{.3\linewidth}}
 \scalebox{.85}{ \parbox{3cm}{
    \begin{fmfgraph}(30,30)
       \fmfset{arrow_len}{2mm}
       \fmfleft{l}
       \fmftop{r1}
       \fmfbottom{r2}
       \fmf{photon,tension=.4}{l,v1}
       \fmf{fermion,left=.5,tension=.2}{v1,v2,v3,v1}
       \fmf{phantom,tension=.1}{v2,r1}
       \fmf{phantom,tension=.1}{v3,r2}
       \fmfright{r}
       \fmf{photon,tension=.3}{r,w1}
       \fmf{fermion,left=.5,tension=.2}{w1,w3,w2,w1}
       \fmf{phantom,tension=.1}{w2,r1}
       \fmf{phantom,tension=.1}{w3,r2}
       \fmf{phantom,tension=.05}{v2,w2}
       \fmf{phantom,tension=.05}{v3,w3}
       \fmfset{wiggly_len}{2mm}
       \fmf{photon,tension=.1}{v2,w3}
       \fmf{photon,tension=.1}{v3,w2}

   \end{fmfgraph}}
 }
 &\scalebox{.85}{ \parbox{3cm}{
    \begin{fmfgraph}(30,30)
       \fmfset{arrow_len}{2mm}
       \fmfleft{l}
       \fmftop{r1}
       \fmfbottom{r2}
       \fmf{photon,tension=.4}{l,v1}
       \fmf{fermion,left=.5,tension=.2}{v1,v2,v3,v1}
       \fmf{phantom,tension=.1}{v2,r1}
       \fmf{phantom,tension=.1}{v3,r2}
       \fmfright{r}
       \fmf{photon,tension=.3}{r,w1}
       \fmf{fermion,left=.5,tension=.2}{w1,w3,w2,w1}
       \fmf{phantom,tension=.1}{w2,r1}
       \fmf{phantom,tension=.1}{w3,r2}
       \fmf{photon,tension=.2}{v2,w2}
       \fmf{photon,tension=.2}{v3,w3}
  \end{fmfgraph}}
 }
 &
 \scalebox{.85}{
  \parbox{2.2cm}{
    \begin{fmfgraph}(30,30)
       \fmfset{arrow_len}{2mm}
       \fmfleft{l1}
       \fmfright{l2}
       \fmftop{r1,r2}
       \fmf{photon,tension=1}{l1,v1}
       \fmf{photon,tension=1}{l2,v2}
       \fmf{fermion,left=.5,tension=2}{v1,v4,v3,v2,v1}
       \fmf{phantom,tension=1}{v4,r1}
       \fmf{phantom,tension=1}{v3,r2}
      \fmf{photon,left,tension=.05}{v4,v3}
 \end{fmfgraph}}}
 \end{tabular}
 
  \caption{The two-point graphs at one-loop.}
  \label{table:1loop-2pt}
\end{table}

\begin{table}
  \begin{tabular}{p{.3\linewidth}p{.3\linewidth}p{.3\linewidth}}
  \scalebox{.85}{ \parbox{2.5cm}{
    \begin{fmfgraph}(30,20)
      \fmfset{arrow_len}{2mm}
             \fmfleft{l1,l,l2}
       \fmftop{r1}
       \fmfbottom{r2}
       \fmf{phantom,tension=.05}{l,v1}
       \fmf{phantom,tension=.05}{l,v1p}
       \fmf{fermion,left=.5,tension=.2}{v1,v1p,v2,v3,v1}
       \fmf{phantom,tension=.1}{v2,r1}
       \fmf{phantom,tension=.1}{v3,r2}
       \fmfright{r}
       \fmf{photon,tension=.3}{r,w1}
       \fmf{fermion,left=.5,tension=.18}{w1,w3,w2,w1}
       \fmf{phantom,tension=.1}{w2,r1}
       \fmf{phantom,tension=.1}{w3,r2}
       \fmf{photon,tension=.2}{v2,w2}
       \fmf{photon,tension=.2}{v3,w3}
       \fmf{photon,tension=.05}{l1,v1}
       \fmf{photon,tension=.05}{l2,v1p}
  \end{fmfgraph}}}
   &   \scalebox{.85}{\parbox{2.2cm}{
    \vspace{-1.5cm}
    \begin{fmfgraph}(30,30)
       \fmfset{arrow_len}{2mm}
       \fmftop{r1,r2}
       \fmfbottom{l1,b,l2}
       \fmf{photon,tension=1}{l1,v1}
       \fmf{photon,tension=1}{l2,v2}
       \fmf{fermion,left=.3,tension=3}{v1,v4,v3,v2,bc,v1}
       \fmf{phantom,tension=1}{v4,r1}
       \fmf{phantom,tension=1}{v3,r2}
       \fmf{photon,left,tension=.01}{v4,v3}
       \fmf{photon}{bc,b}
 \end{fmfgraph}}}
  &
  \scalebox{.85}{\parbox{2.2cm}{
    \begin{fmfgraph}(30,25)
       \fmfset{arrow_len}{2mm}
       \fmftop{l,t,r}
      \fmf{photon,tension=.7}{l,v1}
      \fmf{fermion,left=.5,tension=.3}{v1,v2p,v2,v1}
      \fmf{phantom,tension=.2}{v2p,t}
          \fmfbottom{b}
          \fmf{photon,tension=.7}{b,bc}
            \fmf{photon,tension=.7}{v4,r}
      \fmf{fermion,left=.5,tension=.3}{c2,bc,c1,c2}
      \fmf{photon,tension=.4}{v2,c1}
      \fmf{photon,tension=.4}{c2,v3}
            \fmf{fermion,left=.5,tension=.3}{v3,v4p,v4,v3}
            \fmf{phantom,tension=.2}{v4p,t}
            \fmf{photon,tension=.4}{v2p,v4p}
\end{fmfgraph}}}
 

    \end{tabular}
  \caption{The relevant three-point graphs at one-loop.}
  \label{table:1loop-3pt}
  \end{table}
\subsection{One-loop counterterms to the spectral action}
The computations of the graph amplitudes in the previous section show that the second two graphs in Table \ref{table:1loop-2pt} are the relevant ones to consider as counterterms for the spectral action. However, since the spectral action is in particular a gauge theory, it is crucial that such counterterms are of the same form as the terms appearing in the spectral action.

As may be expected, a crucial role will be played by so-called {\em quantum Ward identities}. They form the analogue of \eqref{eq:ward} for the divergent component of the 1PI $n$-point functions at one loop. Let us denote by $\llangle V_1, \ldots, V_n \rrangle^{1L}$ all one-loop $n$-point graphs whose amplitudes involve a sum over a loop index. The skeletons for such graphs are depicted in Table \ref{table:skel-1l}, for which all external lines are written outside the graph diagram, and labelled in cyclical order. Indeed, if an external line would be in the interior of the diagram, it is surrounded by the loop in the diagram, and will thus prevent the loop index from running (as in Equation \ref{eq:2ptA}).

The quantum Ward identities are now given by
\begin{multline*}
\dbr{V_1,\ldots,aV_j,\ldots,V_n}^{1L}-\dbr{V_1,\ldots,V_{j-1}a,\ldots,V_n}^{1L}\\
        =\dbr{V_1,\ldots,V_{j-1},\diraccom{a},V_j,\ldots,V_n}^{1L}.
\end{multline*}
It is this identity, in combination with cyclicity of the bracket $\llangle V_1, \ldots, V_n \rrangle = \llangle V_n , V_1, \ldots, V_{n-1} \rrangle$, which allows us to follow line-by-line the derivation of the Chern--Simons and Yang--Mills terms in the previous section ({\em cf.} \cite{NS21}). We thus arrive at our main conclusion which is that
   the divergent part of the one-loop quantum effective spectral action  can be expanded as
   $$
\sum_n \frac 1n \dbr{V,\ldots,V}_{\infty}^{1L} 
   = \sum_{k=1}^\infty \left( \int_{\widetilde \psi_{2k-1}}  \!\!\!\!\! \cs_{2k-1} (A) +\frac 1 {2k} \int_{\widetilde \phi_{2k}} \!\!\!\!\! F^{k} \right).
     $$
Here $\widetilde \phi$ and $\widetilde \psi$ are the analogues of $\phi$ and $\psi$ as defined in \eqref{eq:bracket-cochain} and \eqref{eq:psi} but now using the double bracket.
 We conclude that the passage to the one-loop renormalized spectral action can be realized by a transformation in the space of noncommutative integrals, sending $\phi \mapsto \phi - \widetilde \phi$ and $\psi \mapsto \psi- \widetilde \psi$, thus rendering the theory (one-loop) renormalizable as a gauge theory.

\bigskip

Before addressing the general case of $n$-point vertex contributions, we will present a diagrammatic proof of the quantum Ward identity for divergent one-loop two-point functions. 

We first consider the contribution from the second diagram in Table \ref{table:1loop-2pt} to the term $\llangle a V_1, V_2\rrangle - \llangle V_1, V_2 a \rrangle $ in the quantum Ward identity:
\begin{align*}
 &  \scalebox{.7}{ \parbox{3cm}{
    \begin{fmfgraph*}(25,25)
       \fmfset{arrow_len}{2mm}
       \fmfleft{lp,l}
       \fmftop{r1}
       \fmfbottom{r2}
       \fmf{photon,tension=.3,label=$V_1$}{l,v1}
       \fmf{dashes,tension=.3,label=$a$}{v1,lp}
       \fmf{fermion,left=.5,tension=.2}{v1,v2,v3,v1}
       \fmf{phantom,tension=.1}{v2,r1}
       \fmf{phantom,tension=.1}{v3,r2}
       \fmfright{r}
       \fmf{photon,tension=.3,label=$V_2$}{r,w1}
       \fmf{fermion,left=.5,tension=.2}{w1,w3,w2,w1}
       \fmf{phantom,tension=.1}{w2,r1}
       \fmf{phantom,tension=.1}{w3,r2}
       \fmf{photon,tension=.1}{v2,w2}
       \fmf{photon,tension=.1}{v3,w3}
  \end{fmfgraph*}}}
    -\quad 
   \scalebox{.7}{ \parbox{3cm}{
    \begin{fmfgraph*}(25,25)
       \fmfset{arrow_len}{2mm}
       \fmfleft{l}
       \fmftop{r1}
       \fmfbottom{r2}
       \fmf{photon,tension=.3,label=$V_1$}{l,v1}
       \fmf{fermion,left=.5,tension=.2}{v1,v2,v3,v1}
       \fmf{phantom,tension=.1}{v2,r1}
       \fmf{phantom,tension=.1}{v3,r2}
       \fmfright{r,rp}
       \fmf{photon,tension=.3,label=$V_2$}{w1,rp}
       \fmf{dashes,tension=.3,label=$a$}{w1,r}
       \fmf{fermion,left=.5,tension=.2}{w1,w3,w2,w1}
       \fmf{phantom,tension=.1}{w2,r1}
       \fmf{phantom,tension=.1}{w3,r2}
       \fmf{photon,tension=.1}{v2,w2}
       \fmf{photon,tension=.1}{v3,w3}
  \end{fmfgraph*}}}
  \\
  &\quad = \quad \scalebox{.7}{  \parbox{3cm}{
 \blue{    \begin{fmfgraph*}(25,25)
      \fmfset{arrow_len}{2mm}
             \fmfleft{l1,l,l2}
       \fmftop{r1}
       \fmfbottom{r2}
       \fmf{phantom,tension=.05}{l,v1}
       \fmf{phantom,tension=.05}{l,v1p}
       \fmf{fermion,foreground=blue,left=.5,tension=.2}{v1,v1p,v2,v3,v1}
       \fmf{phantom,tension=.1}{v2,r1}
       \fmf{phantom,tension=.1}{v3,r2}
       \fmfright{r}
       \fmf{photon,foreground=blue,tension=.3,label=$V_2$}{r,w1}
       \fmf{fermion,foreground=blue,left=.5,tension=.2}{w1,w3,w2,w1}
       \fmf{phantom,tension=.1}{w2,r1}
       \fmf{phantom,tension=.1}{w3,r2}
       \fmf{photon,foreground=blue,tension=.1}{v2,w2}
       \fmf{photon,foreground=blue,tension=.1}{v3,w3}
       \fmf{photon,foreground=blue,tension=.05,label=$\diraccom{a}$}{v1,l1}
       \fmf{photon,foreground=blue,tension=.05,label=$V_1$}{l2,v1p}
  \end{fmfgraph*}}}}
  + \quad   \scalebox{.7}{  \parbox{3cm}{
   \blue{\begin{fmfgraph*}(25,25)
      \fmfset{arrow_len}{2mm}
             \fmfright{l2,l,l1}
       \fmfbottom{r1}
       \fmftop{r2}
       \fmf{phantom,tension=.05}{l,v1}
       \fmf{phantom,tension=.05}{l,v1p}
       \fmf{fermion,foreground=blue,left=.5,tension=.2}{v1,v1p,v2,v3,v1}
       \fmf{phantom,tension=.1}{v2,r1}
       \fmf{phantom,tension=.1}{v3,r2}
       \fmfleft{r}
       \fmf{photon,foreground=blue,tension=.3,label=$V_1$}{r,w1}
       \fmf{fermion,foreground=blue,left=.5,tension=.2}{w1,w3,w2,w1}
       \fmf{phantom,tension=.1}{w2,r1}
       \fmf{phantom,tension=.1}{w3,r2}
       \fmf{photon,foreground=blue,tension=.1}{v2,w2}
       \fmf{photon,foreground=blue,tension=.1}{v3,w3}
       \fmf{photon,foreground=blue,tension=.05,label=$V_1$}{v1,l1}
       \fmf{photon,foreground=blue,tension=.05,label=$\diraccom{a}$}{l2,v1p}
  \end{fmfgraph*}}}}
  +
    \scalebox{.7}{ \parbox{2.2cm}{
\green{    \begin{fmfgraph*}(25,25)
       \fmfset{arrow_len}{2mm}
      \fmftop{l,t,r}
      \fmf{photon,foreground=green,tension=.7,label=$V_1$}{l,v1}
      \fmf{fermion,foreground=green,left=.5,tension=.3}{v1,v2p,v2,v1}
      \fmf{phantom,tension=.2}{v2p,t}
          \fmfbottom{b}
          \fmf{photon,foreground=green,tension=.7,label=$\diraccom{a}$}{b,bc}
            \fmf{photon,foreground=green,tension=.7,label=$V_2$}{v4,r}
      \fmf{fermion,foreground=green,left=.5,tension=.3}{c2,bc,c1,c2}
      \fmf{photon,foreground=green,tension=.4}{v2,c1}
      \fmf{photon,foreground=green,tension=.4}{c2,v3}
            \fmf{fermion,foreground=green,left=.5,tension=.3}{v3,v4p,v4,v3}
            \fmf{phantom,tension=.2}{v4p,t}
            \fmf{photon,foreground=green,tension=.4}{v2p,v4p}
\end{fmfgraph*}}}}
 \end{align*}

For the third two-point diagram in Table \ref{table:1loop-2pt} there are two possible assignments of the external fields, so that their contribution to $\llangle a V_1, V_2\rrangle - \llangle V_1, V_2 a \rrangle $ is

\begin{align*}
& 
  \scalebox{.7}{ \parbox{2.2cm}{
    \begin{fmfgraph*}(20,20)
       \fmfset{arrow_len}{2mm}
       \fmfleft{l1}
       \fmfright{l2}
       \fmftop{r1,r2}
       \fmf{photon,tension=1,label=$V_1$}{l1,v1}
       \fmf{photon,tension=1,label=$V_2$}{l2,v2}
       \fmf{fermion,left=.5,tension=2}{v1,v4,v3,v2,v1}
       \fmf{phantom,tension=1}{v4,r1}
       \fmf{phantom,tension=1}{v3,r2}
       \fmf{photon,left,tension=.05}{v4,v3}
       \fmffreeze
       \fmfbottom{b}
       \fmf{dashes}{v1,b}
       \fmflabel{$a$}{b}
\end{fmfgraph*}}}  -
  \quad
 \scalebox{.7}{  \parbox{2.5cm}{
    \begin{fmfgraph*}(20,20)
      \fmfset{arrow_len}{2mm}
       \fmfleft{l1}
       \fmfright{l2}
       \fmftop{r1,r2}
       \fmf{photon,tension=1,label=$V_1$}{l1,v1}
       \fmf{photon,tension=1,label=$V_2$}{l2,v2}
       \fmf{fermion,left=.5,tension=2}{v1,v4,v3,v2,v1}
       \fmf{phantom,tension=1}{v4,r1}
       \fmf{phantom,tension=1}{v3,r2}
       \fmf{photon,left,tension=.05}{v4,v3}
       \fmffreeze
       \fmfbottom{b}
       \fmf{dashes}{v2,b}
              \fmflabel{$a$}{b}
 \end{fmfgraph*}}}
=  \scalebox{.7}{ \parbox{3cm}{
    \vspace{-1.5cm}
   \red{ \begin{fmfgraph*}(25,25)
       \fmfset{arrow_len}{2mm}
       \fmftop{r1,r2}
       \fmfbottom{l1,b,l2}
       \fmf{photon,foreground=red,tension=1,label=$V_1$}{l1,v1}
       \fmf{photon,foreground=red,tension=1,label=$V_2$}{l2,v2}
       \fmf{fermion,foreground=red,left=.3,tension=3}{v1,v4,v3,v2,bc,v1}
       \fmf{phantom,tension=1}{v4,r1}
       \fmf{phantom,tension=1}{v3,r2}
       \fmf{photon,foreground=red,left,tension=.01}{v4,v3}
       \fmf{photon,foreground=red}{b,bc}
       \fmflabel{$\diraccom{a}$}{b}
\end{fmfgraph*}}}}
\end{align*}

\medskip

\noindent and
\begin{align*}
&   \scalebox{.7}{ \parbox{3cm}{
    \begin{fmfgraph*}(20,20)
       \fmfset{arrow_len}{2mm}
       \fmfleft{l1}
       \fmfright{l2}
       \fmftop{r1,r2}
       \fmf{photon,tension=1,label=$V_2$}{v1,l1}
       \fmf{photon,tension=1,label=$V_1$}{v2,l2}
       \fmf{fermion,left=.5,tension=2}{v1,v4,v3,v2,v1}
       \fmf{phantom,tension=1}{v4,r1}
       \fmf{phantom,tension=1}{v3,r2}
       \fmf{photon,left,tension=.05}{v4,v3}
       \fmffreeze
       \fmf{dashes}{v2,r2}
       \fmflabel{$a$}{r2}
 \end{fmfgraph*}} }-
\quad    \scalebox{.7}{ \parbox{2.2cm}{
    \begin{fmfgraph*}(20,20)
       \fmfset{arrow_len}{2mm}
       \fmfleft{l1}
       \fmfright{l2}
       \fmftop{r1,r2}
       \fmf{photon,tension=1,label=$V_2$}{v1,l1}
       \fmf{photon,tension=1,label=$V_1$}{v2,l2}
       \fmf{fermion,left=.5,tension=2}{v1,v4,v3,v2,v1}
       \fmf{phantom,tension=1}{v4,r1}
       \fmf{phantom,tension=1}{v3,r2}
       \fmf{photon,left,tension=.05}{v4,v3}
       \fmffreeze
       \fmf{dashes}{v1,r1}
       \fmflabel{$a$}{r1}
\end{fmfgraph*}}}\\
& \quad =  \scalebox{.7}{ \parbox{3cm}{
    \vspace{-1.5cm}
\red{   \begin{fmfgraph*}(25,25)
       \fmfset{arrow_len}{2mm}
       \fmftop{r1,r2}
       \fmfbottom{l1,b,l2}
       \fmf{photon,foreground=red,tension=1,label=$V_2$}{l1,v1}
       \fmf{photon,foreground=red,tension=1,label=$\diraccom{a}$}{l2,v2}
       \fmf{fermion,foreground=red,left=.3,tension=3}{v1,v4,v3,v2,bc,v1}
       \fmf{phantom,tension=1}{v4,r1}
       \fmf{phantom,tension=1}{v3,r2}
       \fmf{photon,foreground=red,left,tension=.01}{v4,v3}
       \fmf{photon,foreground=red,label=$V_1$}{bc,b}
\end{fmfgraph*}}}}
+ \quad \scalebox{.7}{  \parbox{2.7cm}{
    \vspace{-1.5cm}
\red{   \begin{fmfgraph*}(25,25)
       \fmfset{arrow_len}{2mm}
       \fmftop{r1,r2}
       \fmfbottom{l1,b,l2}
       \fmf{photon,foreground=red,tension=1,label=$\diraccom{a}$}{l1,v1}
       \fmf{photon,foreground=red,tension=1,label=$V_1$}{l2,v2}
       \fmf{fermion,foreground=red,left=.3,tension=3}{v1,v4,v3,v2,bc,v1}
       \fmf{phantom,tension=1}{v4,r1}
       \fmf{phantom,tension=1}{v3,r2}
       \fmf{photon,foreground=red,left,tension=.01}{v4,v3}
       \fmf{photon,foreground=red,label=$V_2$}{bc,b}
\end{fmfgraph*}}}}
+    \scalebox{.7}{ \parbox{3cm}{
   \blue{ \begin{fmfgraph*}(35,20)
      \fmfset{arrow_len}{2mm}
             \fmfleft{l1,l,l2}
       \fmftop{r1}
       \fmfbottom{r2}
       \fmf{phantom,tension=.05}{l,v1}
       \fmf{phantom,tension=.05}{l,v1p}
       \fmf{fermion,foreground=blue,left=.5,tension=.2}{v1,v1p,v2,v3,v1}
       \fmf{phantom,tension=.1}{v2,r1}
       \fmf{phantom,tension=.1}{v3,r2}
       \fmfright{r}
       \fmf{photon,foreground=blue,tension=.3,label=$\diraccom{a}$}{r,w1}
       \fmf{fermion,foreground=blue,left=.5,tension=.18}{w1,w3,w2,w1}
       \fmf{phantom,tension=.1}{w2,r1}
       \fmf{phantom,tension=.1}{w3,r2}
       \fmf{photon,foreground=blue,tension=.2}{v2,w2}
       \fmf{photon,foreground=blue,tension=.2}{v3,w3}
       \fmf{photon,foreground=blue,tension=.05,label=$V_1$}{l1,v1}
       \fmf{photon,foreground=blue,tension=.05,label=$V_2$}{v1p,l2}
  \end{fmfgraph*}}}}
\end{align*}
We have coloured the Feynman graphs on the right-hand side of the quantum Ward identity according to their topology, {\em i.e.} as they appear in Table \ref{table:1loop-3pt}. One then readily sees that the graphs conspire to yield all cyclic permutations of $\diraccom{a},V_1,V_2$ as external fields on all planar one-loop graphs with three external legs. 

\medskip

\begin{table}
\centering
 \begin{tabular}{p{.22\textwidth}p{.22\textwidth}p{.22\textwidth}p{.22\textwidth}}
   \scalebox{.7}{\parbox{6cm}{
        
    \begin{fmfgraph}(30,30)
      \fmfbottom{l1,l2,l3}
      \fmftop{r}
      \fmf{photon}{l2,v1}
      \fmf{photon,left}{v1,r,v1}
      \fmf{photon}{l1,v1}
      \fmf{photon}{l3,v1}
      \fmfv{decor.shape=circle,decor.filled=empty,decor.size=.15w}{v1}
      \fmf{phantom,right=.4}{l1,l3}
      \fmfipath{p}
      \fmfiset{p}{vpath(__l1,__l3)}
      \fmfi{dots}{subpath (length(p)/6,5*length(p)/6) of p}
    \end{fmfgraph}}}&
    
    \scalebox{.7}{\parbox{6cm}{
   
    \begin{fmfgraph}(30,30)
      \fmfleft{l1,l2,l3}
      \fmfright{r1,r2,r3}
      \fmf{photon}{l2,v1}
      \fmf{photon}{r2,v2}
      \fmf{photon,left,tension=.5}{v1,v2,v1}
      \fmffreeze
      \fmf{photon}{l1,v1}
      \fmf{photon}{r1,v2}
      \fmf{photon}{l3,v1}
      \fmf{photon}{r3,v2}
      \fmfv{decor.shape=circle,decor.filled=empty,decor.size=.15w}{v1}
      \fmfv{decor.shape=circle,decor.filled=empty,decor.size=.15w}{v2}
           \fmf{phantom,left=.4}{l1,l3}
      \fmfipath{p[]}
      \fmfiset{p1}{vpath(__l1,__l3)}
      \fmfi{dots}{subpath (length(p1)/6,5*length(p1)/6) of p1}
           \fmf{phantom,right=.4}{r1,r3}
      \fmfiset{p2}{vpath(__r1,__r3)}
      \fmfi{dots}{subpath (length(p2)/6,5*length(p2)/6) of p2}
%
    \end{fmfgraph}}}
	&    
    
 \scalebox{.7}{\parbox{6cm}{
    
    \begin{fmfgraph}(30,30)
      \fmfleft{l2,l3}
      \fmfright{r2,r3}
      \fmftop{t1,t2,t3}
      \fmfbottom{b1,b3}
      \fmf{photon}{l2,v1}
      \fmf{photon}{r2,v3}
      \fmf{photon}{t2,v2}
      \fmf{photon,tension=.5}{v1,v2,v3,v1}
            \fmffreeze
      \fmf{photon}{l3,v2}
      \fmf{phantom}{t1,v1}
      \fmf{phantom}{t3,v3}
      \fmf{photon}{r3,v2}
      \fmf{photon}{b1,v1}
      \fmf{photon}{b3,v3}
      \fmfv{decor.shape=circle,decor.filled=empty,decor.size=.15w}{v1}
      \fmfv{decor.shape=circle,decor.filled=empty,decor.size=.15w}{v2}
      \fmfv{decor.shape=circle,decor.filled=empty,decor.size=.15w}{v3}
      \fmfipath{p[]}
      \fmf{phantom,left=.4}{l3,r3}
      \fmfiset{p1}{vpath(__l3,__r3)}
      \fmfi{dots}{subpath (length(p1)/6,5*length(p1)/6) of p1}
       \fmf{phantom,right=.4}{b1,l2}
      \fmfiset{p2}{vpath(__b1,__l2)}
      \fmfi{dots}{subpath (length(p2)/6,5*length(p2)/6) of p2}
       \fmf{phantom,left=.4}{b3,r2}
      \fmfiset{p3}{vpath(__b3,__r2)}
      \fmfi{dots}{subpath (length(p3)/6,5*length(p1)/6) of p3}
\end{fmfgraph}}}
  &
 \scalebox{.7}{\parbox{6cm}{
 
    \begin{fmfgraph}(30,30)
      \fmfleft{l1,l2,l3}
      \fmfright{r1,r2,r3}
      \fmftop{t1,t2,t3}
      \fmfbottom{b1,b2,b3}
      \fmf{photon}{l2,v1}
      \fmf{photon}{r2,v3}
      \fmf{photon}{t2,v2}
      \fmf{photon}{b2,v4}
      \fmf{photon,tension=.5}{v1,v2,v3,v4,v1}
            \fmffreeze
      \fmf{photon}{l1,v4}
      \fmf{photon}{l3,v2}
      \fmf{photon}{t1,v1}
      \fmf{photon}{t3,v3}
      \fmf{photon}{r1,v4}
      \fmf{photon}{r3,v2}
      \fmf{photon}{b1,v1}
      \fmf{photon}{b3,v3}
      \fmfv{decor.shape=circle,decor.filled=empty,decor.size=.15w}{v1}
      \fmfv{decor.shape=circle,decor.filled=empty,decor.size=.15w}{v2}
      \fmfv{decor.shape=circle,decor.filled=empty,decor.size=.15w}{v3}
      \fmfv{decor.shape=circle,decor.filled=empty,decor.size=.15w}{v4}
      \fmfipath{p[]}
      \fmf{phantom,left=.4}{l3,r3}
      \fmfiset{p1}{vpath(__l3,__r3)}
      \fmfi{dots}{subpath (length(p1)/6,5*length(p1)/6) of p1}
      \fmf{phantom,right=.4}{t1,b1}
      \fmfiset{p2}{vpath(__t1,__b1)}
      \fmfi{dots}{subpath (length(p2)/6,5*length(p2)/6) of p2}
      \fmf{phantom,left=.4}{t3,b3}
      \fmfiset{p3}{vpath(__t3,__b3)}
      \fmfi{dots}{subpath (length(p3)/6,5*length(p3)/6) of p3}
      \fmf{phantom,right=.4}{l1,r1}
      \fmfiset{p4}{vpath(__l1,__r1)}
      \fmfi{dots}{subpath (length(p4)/6,5*length(p4)/6) of p4}
 \end{fmfgraph}}}
 \end{tabular}
 \caption{Skeletons for divergent one-loop $n$-point functions with increasing number of vertices. The fermion loops that define the vertices are all oriented as clockwise.}
 \label{table:skel-1l}
\end{table}

This argument extends to all potentially divergent one-loop $n$-point functions $\llangle V_1, \ldots, V_n \rrangle^{1L}$ as follows. Recall that all such divergent one-loop diagrams have skeletons as depicted in Table \ref{table:skel-1l}, with the external lines labelled cyclically from $1$ to $n$.
The decoration of the external legs of our graphs with the external fields $V_1, \ldots, V_n$ then proceeds according to this labelling $1, \ldots, n$ and, upon summing over all such decorated graphs $G$, we get
$$
\llangle V_1, \ldots, V_n \rrangle^{1L} = \sum_G 
G_{ V_{1},\ldots V_{n} }.
$$

The left-hand side of the quantum Ward identity essentially comes down to connecting external edges to the graphs $G$. We will write $G_i$ for the graph $G$ with an insertion of an external gauge edge at a point $i$ in between $n$ and $1$: this insertion point $i$ can be either an outer fermion line in $G$ (as in \eqref{eq:ward}) or, if $1$ and $n$ are not attached to the same vertex in $G$, a gauge propagator (as in \eqref{eq:ward-gauge}). We then find 
\begin{align*}
\llangle a  V_1, \ldots, V_n \rrangle^{1L} - 
\llangle V_1, \ldots, V_n  a \rrangle^{1L} & = \sum_{G,i} 
(G_i)_{ \diraccom{a} , V_1, \ldots ,V_n  },
\end{align*}
where the decoration $\diraccom{a}$ is attached to the external gauge edge inserted at the point $i$ of $G_i$.

It is clear that the sum over $G$ and $i$ yield all decorated $n+1$-point graphs, and, moreover, that any $n+1$-point graph with labels $\diraccom{a} , V_1, \ldots ,V_n$ is obtained in a unique manner from an insertion of an external edge in an $n$-point graph, as described above. We are thus left precisely with $\llangle {\diraccom{a}},  V_1, \ldots, V_n \rrangle^{1L}$ as desired.

\section{Conclusions}
In this paper we have analyzed the quantum gauge fluctuations for the spectral action in noncommutative geometry. Using the background field method we have showed one-loop renormalizability of the spectral action, while staying within the same spectral framework.

Naturally, this forms the starting point for more direct applications of noncommutative geometry to particle physics phenomenology. Instead of the spectral action playing the role of a bare action functional, to which subsequent RG-methods are applied, we now have a candidate for a so-called {\em quantum effective spectral action}, given by the sum of all 1PI Feynman diagrams and which is supposed to be valid at all energies. 
One may then try to extend the derivation of bare physical Lagrangians from the spectral action \cite{CC96,Sui14} to the renormalized spectral action, and arrive at a spectral, noncommutative geometric description of particle physics which is also valid and falsifiable at lower energies.

Besides these future steps in the applications to particle physics phenomenology, it is also important to extend the ``power-counting'' and diagrammatics of the one-loop renormalizability that we presented here to arbitrary loop order. This, and also a more detailed account of the derivation presented in this paper, will be reported elsewhere. 
The connection with the proof of renormalizability for noncommutative scalar field theories \cite{GW05} also deserves further investigation. One of the main differences is that they consider so-called non-local matrix models \cite{GW05b} with a quartic vertex, while instead we have a local matrix model but with vertices of arbitrary valence.

  \subsection*{Acknowledgements}
  We thank Steven Lord, Fedor Sukochev and the other members in their research group for fruitful discussions during a visit in August 2019, where the basis for the current paper was laid. We also thank Ali Chamseddine and Alain Connes for useful comments.

  Research supported by NWO Physics Projectruimte (680-91-101).

\newcommand{\noopsort}[1]{}\def\cprime{$'$}

\end{fmffile}

\begin{thebibliography}{10}

\bibitem{ASZ15}
N.~Alkofer, F.~Saueressig, and O.~Zanusso.
\newblock {Spectral dimensions from the spectral action}.
\newblock {\em Phys. Rev.} D91 (2015)  025025.

\bibitem{AK19}
S.~Azarfar and M.~Khalkhali.
\newblock Random finite noncommutative geometries and topological recursion,
  1906.09362.

\bibitem{BG16}
J.~W. Barrett and L.~Glaser.
\newblock {Monte Carlo simulations of random non-commutative geometries}.
\newblock {\em J. Phys.} A49 (2016)  245001.

\bibitem{BBS16}
W.~Beenakker, T.~van~den Broek, and W.~D. van Suijlekom.
\newblock {\em Supersymmetry and noncommutative geometry}, volume~9 of {\em
  SpringerBriefs in Mathematical Physics}.
\newblock Springer, Cham, 2016.

\bibitem{BIZ80}
D.~Bessis, C.~Itzykson, and J.~B. Zuber.
\newblock {Quantum field theory techniques in graphical enumeration}.
\newblock {\em Adv. Appl. Math.} 1 (1980)  109--157.

\bibitem{BS20b}
A.~Bochniak and A.~Sitarz.
\newblock {Spectral geometry for the standard model without fermion doubling}.
\newblock {\em Phys. Rev. D} 101 (2020)  075038.

\bibitem{BF14}
L.~Boyle and S.~Farnsworth.
\newblock {Non-Commutative Geometry, Non-Associative Geometry and the Standard
  Model of Particle Physics}.
\newblock {\em New J. Phys.} 16 (2014)  123027.

\bibitem{BroS11}
T.~{\noopsort{broek}}van~den Broek and W.~D. van Suijlekom.
\newblock {Supersymmetric QCD from noncommutative geometry}.
\newblock {\em Phys. Lett.} B699 (2011)  119--122.

\bibitem{CC96}
A.~H. Chamseddine and A.~Connes.
\newblock Universal formula for noncommutative geometry actions: {U}nifications
  of gravity and the {S}tandard {M}odel.
\newblock {\em Phys. Rev. Lett.} 77 (1996)  4868--4871.

\bibitem{CCM07}
A.~H. Chamseddine, A.~Connes, and M.~Marcolli.
\newblock Gravity and the {S}tandard {M}odel with neutrino mixing.
\newblock {\em Adv. Theor. Math. Phys.} 11 (2007)  991--1089.

\bibitem{CCM15}
A.~H. Chamseddine, A.~Connes, and V.~Mukhanov.
\newblock Quanta of geometry: {N}oncommutative aspects.
\newblock {\em Phys. Rev. Lett.} 114 (2015)  091302.

\bibitem{CCS13b}
A.~H. Chamseddine, A.~Connes, and W.~D. van Suijlekom.
\newblock {Beyond the spectral Standard Model: Emergence of Pati-Salam
  unification}.
\newblock {\em JHEP} 1311 (2013)  132.

\bibitem{CCS15}
A.~H. Chamseddine, A.~Connes, and W.~D. van Suijlekom.
\newblock Grand unification in the spectral {P}ati-{S}alam model.
\newblock {\em JHEP} 11 (2015)  011.

\bibitem{CIS20}
A.~H. Chamseddine, J.~Iliopoulos, and W.~D. van Suijlekom.
\newblock {Spectral action in matrix form}.
\newblock {\em Eur. Phys. J. C} 80 (2020)  1045.

\bibitem{C94}
A.~Connes.
\newblock {\em Noncommutative Geometry}.
\newblock Academic Press, San Diego, 1994.

\bibitem{C96}
A.~Connes.
\newblock Gravity coupled with matter and the foundation of non-commutative
  geometry.
\newblock {\em Commun. Math. Phys.} 182 (1996)  155--176.

\bibitem{CM07}
A.~Connes and M.~Marcolli.
\newblock {\em Noncommutative Geometry, Quantum Fields and Motives}.
\newblock AMS, Providence, 2008.

\bibitem{CC06}
A.~Connes and A.~H. Chamseddine.
\newblock {Inner fluctuations of the spectral action}.
\newblock {\em J. Geom. Phys.} 57 (2006)  1--21.

\bibitem{DAS18}
L.~Dabrowski, F.~D'Andrea, and A.~Sitarz.
\newblock The {S}tandard {M}odel in noncommutative geometry: fundamental
  fermions as internal forms.
\newblock {\em Lett. Math. Phys.} 108 (2018)  1323--1340.

\bibitem{DS18}
L.~Dabrowski and A.~Sitarz.
\newblock {Fermion masses, mass-mixing and the almost commutative geometry of
  the Standard Model}.
\newblock {\em JHEP} 02 (2019)  068.

\bibitem{DLM14b}
A.~Devastato, F.~Lizzi, and P.~Martinetti.
\newblock Higgs mass in noncommutative geometry.
\newblock {\em Fortschr. Phys.} 62 (2014)  863--868.

\bibitem{DLM14}
A.~Devastato, F.~Lizzi, and P.~Martinetti.
\newblock Grand symmetry, spectral action, and the {H}iggs mass.
\newblock {\em JHEP} 1401 (2014)  042.

\bibitem{DM14}
A.~Devastato and P.~Martinetti.
\newblock {Twisted spectral triple for the Standard Model and spontaneous
  breaking of the Grand Symmetry}.
\newblock {\em Math. Phys. Anal. Geom.} 20 (2017) ~2.

\bibitem{GS19a}
L.~Glaser and A.~Stern.
\newblock {Understanding truncated non-commutative geometries through computer
  simulations}.
\newblock {\em J. Math. Phys.} 61 (2020)  033507.

\bibitem{GS19b}
L.~Glaser and A.~B. Stern.
\newblock Reconstructing manifolds from truncations of spectral triples.
\newblock {\em J. Geom. Phys.} 159 (2021)  Paper No. 103921, 17.

\bibitem{GW96}
J.~Gomis and S.~Weinberg.
\newblock Are nonrenormalizable gauge theories renormalizable?
\newblock {\em Nucl. Phys.} B469 (1996)  473--487.

\bibitem{GW05b}
H.~Grosse and R.~Wulkenhaar.
\newblock Power-counting theorem for non-local matrix models and
  renormalisation.
\newblock {\em Comm. Math. Phys.} 254 (2005)  91--127.

\bibitem{GW05}
H.~Grosse and R.~Wulkenhaar.
\newblock {Renormalization of $\phi^4$ theory on noncommutative $\mathbb R^4$
  in the matrix base}.
\newblock {\em Commun. Math. Phys.} 256 (2005)  305--374.

\bibitem{ILV11}
B.~Iochum, C.~Levy, and D.~Vassilevich.
\newblock {Spectral action beyond the weak-field approximation}.
\newblock {\em Commun. Math. Phys.} 316 (2012)  595--613.

\bibitem{Ise19}
R.~A. Iseppi.
\newblock The {BV} formalism: theory and application to a matrix model.
\newblock {\em Rev. Math. Phys.} 31 (2019)  1950035, 24.

\bibitem{IS17}
R.~A. Iseppi and W.~D. van Suijlekom.
\newblock Noncommutative geometry and the {BV} formalism: application to a
  matrix model.
\newblock {\em J. Geom. Phys.} 120 (2017)  129--141.

\bibitem{KP20}
M.~Khalkhali and N.~Pagliaroli.
\newblock Phase transition in random noncommutative geometries.
\newblock {\em Journal of Physics A: Mathematical and Theoretical} 54 (2020)
  035202.

\bibitem{KLV13}
M.~A. Kurkov, F.~Lizzi, and D.~Vassilevich.
\newblock {High energy bosons do not propagate}.
\newblock {\em Phys. Lett. B} 731 (2014)  311--315.

\bibitem{Nak90}
M.~Nakahara.
\newblock {\em Geometry, Topology and Physics}.
\newblock IOP Publishing, 1990.

\bibitem{BS20}
A.~Sitarz.
\newblock {Towards the signs of new physics through the spectral action}.
\newblock {\em Int. J. Geom. Meth. Mod. Phys.} 17 (2020)  2040008.

\bibitem{Skr13}
A.~Skripka.
\newblock Asymptotic expansions for trace functionals.
\newblock {\em J. Funct. Anal.} 266 (2014)  2845--2866.

\bibitem{Sui11}
W.~D. {\noopsort{suijlekom}}van Suijlekom.
\newblock Perturbations and operator trace functions.
\newblock {\em J. Funct. Anal.} 260 (2011)  2483--2496.

\bibitem{Sui11b}
W.~D. {\noopsort{suijlekom}}van Suijlekom.
\newblock {Renormalization of the spectral action for the Yang-Mills system}.
\newblock {\em JHEP} 1103 (2011)  146.

\bibitem{Sui14}
W.~D. {\noopsort{suijlekom}}van Suijlekom.
\newblock {\em Noncommutative Geometry and Particle Physics}.
\newblock Springer, 2015.

\bibitem{NS21}
T.~D.~H. van Nuland and W.~D. van Suijlekom.
\newblock Cyclic cocycles in the spectral action.
\newblock {\em J. Noncommut. Geom.}, to appear (arXiv:2104.09899).

\end{thebibliography}
\end{document}